Real-time absolute frequency measurement of continuous-wave terahertz wave based on dual terahertz combs of photocarriers with different frequency spacings

Takeshi Yasui,[1,2,3] Kenta Hayashi,[1] Ryuji Ichikawa,[1] Harsono Cahyadi,[1,3] Yi-Da Hsieh,[1,3] Yasuhiro Mizutani,[1,3] Hirotsugu Yamamoto,[3,4] Tetsuo Iwata,[1,3] Hajime Inaba,[3,5] and Kaoru Minoshima[3,6]

[1]Institute of Technology and Science, Tokushima University, 2-1 Minami-Josanjima, Tokushima 770-8506, Japan

[2]Graduate School of Engineering Science, Osaka University, 1-3 Machikaneyama, Toyonaka, Osaka 560-8531, Japan

[3]JST, ERATO, MINOSHIMA Intelligent Optical Synthesizer Project, 2-1 Minami-Josanjima, Tokushima 770-8506, Japan

[4]Center for Optical Research and Education, Utsunomiya University, 7-1-2, Yoto, Utsunomiya, Tochigi 321-858, Japan

[5]National Metrology Institute of Japan, National Institute of Advanced Industrial Science and Technology, 1-1-1 Umezono, Tsukuba, Ibaraki 305-8563, Japan

[6]Graduate School of Informatics and Engineering, The University of Electro-Communications, 1-5-1, Chofugaoka, Chofu, Tokyo 182-8585, Japan

# Abstract

Real-time measurement of the absolute frequency of continuous-wave terahertz (CW-THz) waves is required for characterization and frequency calibration of practical CW-THz sources. We proposed a method for real-time monitoring of the absolute frequency of CW-THz waves involving temporally parallel, i.e., simultaneous, measurement of two pairs of beat frequencies and laser repetition frequencies based on dual THz combs of photocarriers (PC-THz combs) with different frequency spacings. To demonstrate the method, THz-comb-referenced spectrum analyzers were constructed with a dual configuration based on dual femtosecond lasers. Regardless of the presence or absence of frequency control in the PC-THz combs, a frequency precision of $10^{-11}$ was achieved at a measurement rate of 100 Hz. Furthermore, large fluctuation of the CW-THz frequencies, crossing several modes of the PC-THz combs, was correctly monitored in real time. The proposed method will be a powerful tool for the research and development of practical CW-THz sources, and other applications.

# 1. Introduction

When femtosecond mode-locked laser light is incident onto a photoconductive antenna (PCA) for detecting terahertz (THz) waves, sub-picosecond photoconductive switching is repeated in the PCA in synchronization with the laser pulses. The sequence of switching operations is essentially copies of the same switching operation separated by an interval equal to the laser repetition period. This highly stable, switching pulse train in the time domain can be synthesized by a series of frequency spikes of photocarrier generation regularly separated by the laser repetition frequency in the frequency domain [1]. This structure is referred to as a THz frequency comb of photocarriers, or a PC-THz comb. Since the absolute frequencies of all frequency modes in the PC-THz comb can be phase-locked to a microwave frequency standard by control of the laser repetition frequency, such a frequency-comb structure enables us to use a PC-THz comb as a precise ruler for measuring THz frequency.

Recently, the potential of PC-THz combs in THz frequency metrology has been recognized [2, 3], for example, as a THz-comb-referenced spectrum analyzer or frequency counter for absolute frequency measurement [4–7]. This type of spectrum analyzer is capable of precise frequency measurement within the frequency coverage of the PC-THz comb at room temperature by using the following procedure: First, a PC-THz comb is generated in a PCA. Second, a continuous-wave THz (CW-THz) wave is mixed with the generated PC-THz comb. Finally, the resultant signal is beat-down to the radio-frequency (RF) region by photoconductive mixing. A THz-comb-referenced spectrum analyzer based on this technique has been successively applied to the absolute frequency measurement of narrow-linewidth CW-THz waves [4–7] and even broadband THz combs [1, 8–10]. Similar approaches for CW-THz waves have been demonstrated in combination with free-space electro-optics sampling [11, 12] and an interferometric method [13] in place of the photoconductive detection. Also, PC-THz combs have been used in the phase and its slope measurements of tunable CW-THz waves for THz distance measurement of optically rough objects [14]. Furthermore, the generation of a frequency standard signal has been achieved by using a PC-THz comb in combination with frequency control of the CW-THz sources [15, 16].

In previous studies on THz-comb-referenced spectrum analyzers, a single PC-THz comb has been used [4–6]. In this case, it has been necessary to measure two beat frequencies respectively corresponding to two different frequency spacings of the PC-THz comb in order to determine the comb mode number nearest in frequency to the CW-THz wave. Therefore, two beat frequencies have been measured before and after shifting the frequency spacing of the PC-THz comb by the laser control. This temporally serial, two-step measurement with a single PC-THz comb has been an obstacle in applying this technique to the real-time absolute frequency measurement of frequency-fluctuating CW-THz waves. Also, use of a precisely stabilized femtosecond laser comb often hinders the easy use of this spectrum analyzer. If the real-time absolute frequency measurement of practical CW-THz sources with rapid, large frequency variations could be implemented using unstabilized femtosecond lasers, the scope of applications would be greatly expanded.

In the work described in this article, we determined the absolute frequency of a frequency-fluctuating CW-THz wave in real time based on temporally parallel, i.e. simultaneous, measurement of two pairs of beat frequencies and repetition frequencies of dual PC-THz combs with different frequency spacings, that is to say, by using a dual THz-comb-referenced spectrum analyzer. We also investigated the possibility of using a PC-THz comb without frequency stabilization for the real-time absolute frequency measurement of the frequency-fluctuating CW-THz wave.

## 2. Principle

THz-comb-referenced spectrum analyzer is based on a heterodyne technique involving photoconductive mixing between a PC-THz comb and a CW-THz wave, which is described in detail elsewhere [4, 5]. There are two essential points in this method: First, a PCA is used as a heterodyne receiver having high, broadband spectral sensitivity in the THz region without the need for cryogenic cooling. Second, the PC-THz comb functions as a local oscillator with multiple frequencies, fully covering the THz region.

In the photoconductive mixing, the absolute frequency of the measured

CW-THz wave (= $f_{THz}$) is given by

$$f_{THz} = mf_{rep} \pm f_{beat}, \tag{1}$$

where $m$ is the order of the comb mode nearest in frequency to the CW-THz wave, $f_{rep}$ is the repetition frequency of the femtosecond laser, and $f_{beat}$ is the lowest frequency of the beat signals. Since $f_{rep}$ and $f_{beat}$ can be measured directly in the RF region, the value of $m$ and the sign of $f_{beat}$ have to be determined to obtain $f_{THz}$. To this end, one has to measure two $f_{beat}$ values ($f_{beat1}$ and $f_{beat2}$) corresponding to two different $f_{rep}$ values ($f_{rep1}$ and $f_{rep2}$), because the relation between them is given by

$$-m(f_{rep2} - f_{rep1}) = f_{beat2} - f_{beat1}. \tag{2}$$

Since previous studies have been based on a single PC-THz comb, it is essential to measure the beat frequencies ($f_{beat1}$ and $f_{beat2}$) before and after shifting the frequency spacing of the PC-THz comb by laser control ($f_{rep1}$ and $f_{rep2}$) [4-6]. However, such temporally serial, two-step measurement with a single PC-THz comb hinders the real-time determination of $f_{THz}$. For real-time determination, temporally parallel, that is, simultaneous, measurement of $f_{beat1}$, $f_{beat2}$, $f_{rep1}$, and $f_{rep2}$ should be performed. To this end, the use of dual PC-THz combs with different frequency spacings will be useful.

Figure 1 shows the principle of real-time determination of $f_{THz}$ based on dual PC-THz combs with different frequency spacings. When two femtosecond lasers with $f_{rep1}$ and $f_{rep2}$ are incident onto two different PCAs (PCA1 and PCA2), two PC-THz combs (PC-THz-comb1 and PC-THz-comb2) with slightly detuned frequency spacings ($f_{rep1}$ and $f_{rep2}$) are respectively induced in them. Then, a measured CW-THz wave (frequency $f_{THz}$) is incident on both PCA1 and PCA2, resulting in the generation of two beat signals with frequencies $f_{beat1}$ and $f_{beat2}$. From Eq. (2), the $m$ value can be obtained by

$$m = \frac{|f_{beat2} - f_{beat1}|}{|f_{rep2} - f_{rep1}|} \quad . \tag{3}$$

Finally, $f_{THz}$ can be determined by measuring $f_{rep1}$, $f_{rep2}$, $f_{beat1}$, and $f_{beat2}$ because

$$\begin{aligned} f_{THz} &= mf_{rep1} + f_{beat1} = \frac{|f_{beat2} - f_{beat1}|}{|f_{rep2} - f_{rep1}|} f_{rep1} + f_{beat1} \quad & \left\langle \frac{f_{beat2} - f_{beat1}}{f_{rep2} - f_{rep1}} < 0 \right\rangle \\ f_{THz} &= mf_{rep1} - f_{beat1} = \frac{|f_{beat2} - f_{beat1}|}{|f_{rep2} - f_{rep1}|} f_{rep1} - f_{beat1} \quad & \left\langle \frac{f_{beat2} - f_{beat1}}{f_{rep2} - f_{rep1}} > 0 \right\rangle \end{aligned} \quad . \tag{4}$$

If $f_{rep1}$ and $f_{rep2}$ are stabilized at known values by laser control, we need to measure just $f_{beat1}$ and $f_{beat2}$ to determine $f_{THz}$. If $f_{rep1}$ and $f_{rep2}$ are fluctuated due to the free-running operation, $f_{beat1}$, $f_{beat2}$, $f_{rep1}$, and $f_{rep2}$ should be measured at the same time.

## 3. Experimental setup

The THz-comb-referenced spectrum analyzer that we developed was composed of femtosecond lasers, a PCA for THz detection, and data acquisition electronics. Two of these THz-comb-referenced spectrum analyzers were constructed with a dual configuration based on dual femtosecond lasers, and these were effectively used to determine the absolute frequency of a CW-THz wave in real time. Figure 2 shows a schematic diagram of the experimental setup. We used dual mode-locked Er-doped fiber lasers (ASOPS TWIN 100 with P100, Menlo Systems; center wavelength = 1550 nm, pulse duration = 50 fs) with slightly mismatched repetition frequencies ($f_{rep1}$ and $f_{rep2}$) to generate dual PC-THz combs with different frequency spacings in the PCAs. When the repetition-frequency stabilization systems were activated, $f_{rep1}$ and $f_{rep2}$ were phase-locked to a rubidium frequency standard (Stanford Research Systems FS725 with frequency = 10 MHz, accuracy = $5\times10^{-11}$, stability = $2\times10^{-11}$ at 1 s).

The two laser-beams were individually focused on bowtie-shaped, low-temperature-grown GaAs photoconductive antennas (PCA1 and PCA2) after wavelength conversion by second harmonic generation (SHG) with periodically-poled-lithium-niobate (PPLN) crystals. Dual PC-THz combs with frequency spacings of $f_{rep1}$ and $f_{rep2}$ (PC-THz-comb1 and PC-THz-comb2) were respectively induced in PCA1 and PCA2. To test the spectrum analyzers, we measured a CW-THz wave (frequency $f_{THz}$) from an active frequency multiplier chain (Millitech AMC-10-R0000, multiplication factor = 6, tuning range = 75–110 GHz, linewidth <0.6 Hz, and average power = 2.5 mW), which multiplied the output frequency of a microwave frequency synthesizer (Agilent E8257D, frequency = 12.5–18.33 GHz, and linewidth <0.1 Hz) by six. The frequency synthesizer is phase-locked to the frequency standard. When this CW-THz wave was incident on both PCA1 and PCA2 together with the two laser beams, groups of current beat signals between the CW-THz wave and PC-THz-comb1/PC-THz-comb2 were generated in the RF region as a result of photoconductive mixing. The beat signals at the lowest frequency (frequencies $f_{beat1}$ and $f_{beat2}$) were extracted by amplifying and low-pass-filtering with current preamplifiers (AMP; bandwidth = 40 MHz and sensitivity = 100,000 V/A). The signals with frequencies $f_{rep1}$ and $f_{rep2}$ were measured by detecting the laser beams with fast photodiodes. The four temporal waveforms for $f_{beat1}$, $f_{beat2}$, $f_{rep1}$, and $f_{rep2}$ were acquired simultaneously by a fast digitizer (resolution = 14 bit, sampling rate = 100 MHz). We applied a software-based, instantaneous-frequency-calculation algorithm to the acquired temporal waveforms [6]. This algorithm is consisting of Fourier transform, digital frequency filtering, inverse Fourier transform, Hilbert transform, and time differential of instantaneous phase. Finally, we determined $f_{THz}$ by substituting the instantaneous values of $f_{beat1}$, $f_{beat2}$, $f_{rep1}$, and $f_{rep2}$ into Eq. (4).

## 4. Results

### 4.1 Use of dual PC-THz combs with stabilization of frequency spacing

First, we demonstrated real-time monitoring of $f_{THz}$ for the frequency-fixed CW-THz wave based on dual PC-THz combs with stabilization of $f_{rep1}$ and $f_{rep2}$, that is to say, using stabilized dual PC-THz combs. Figures 3(a) and (b) show temporal changes of instantaneous values of $f_{beat1}$ and $f_{beat2}$ when $f_{rep1}$ and $f_{rep2}$ were fixed at 100,000,000 Hz and 100,000,050 Hz by laser stabilization control. From these values, we obtained the temporal change of $f_{THz}$ as shown in Fig. 3(c). The mean and standard deviation of $f_{THz}$ were 100,001,000,040 Hz and 60 Hz, respectively. Since $f_{THz}$ was stabilized at 100,001,000,040 THz within a frequency range of 1 Hz by providing the frequency signal from the frequency standard to the frequency synthesizer as an external reference, the variation of $f_{THz}$ in Fig. 3(c) was mainly due to the error in the instantaneous-frequency-calculation algorithm of $f_{beat1}$ and $f_{beat2}$.

To reduce the errors in $f_{THz}$, we performed signal averaging of $f_{beat1}$ after calculating its instantaneous values. Figure 4(a) shows the relationship between the number of averaged signals and the frequency error in $f_{beat1}$, and the upper horizontal axis gives the corresponding measurement rate. It is clear that the frequency error depends on the number of averaged signals or the measurement rate. To avoid the incorrect determination of $m$, we have to determine $f_{beat1}$ and $f_{beat2}$ within a frequency error of ± 25 Hz because $|f_{rep2} - f_{rep1}|$ in Eq. (3) was set to 50 Hz in this experiment. Therefore, we need to average at least 1,000 signals, corresponding to a measurement rate of 10 kHz, to determine $m$ correctly. We also investigated the frequency precision of $f_{THz}$ and the corresponding frequency error with respect to the signal-to-noise ratio (SNR) when the measurement rate was set to 10 Hz, as shown in Fig. 4(b). For comparison, we also show the relation between them for frequency measurement with an RF frequency counter, which has been used in previous studies [4, 5, 7]. Although both methods showed the dependence of the frequency precision on the SNR, their dependence characteristics were different from each other. An SNR of only 10 dB was sufficient for achieving a frequency precision of $10^{-11}$ in the instantaneous-frequency-calculation algorithm, whereas the frequency counter method required an SNR of at least 30 dB to perform the absolute frequency measurement due to the measurement principle of the RF frequency counter. Thus, a combination of dual PC-THz combs with the instantaneous-frequency-calculation algorithm enables precise measurement of $f_{THz}$ in real time and/or at low SNR. We achieved the frequency precision of $8.2 \times 10^{-12}$ at a measurement rate of 100 Hz. It should be note that this value indicates the relative precision of frequency measurement because the output signal of the frequency standard is used as a common time-base signal in this setup. The absolute precision of frequency measurement is determined by the accuracy of the frequency standard or the relative precision whichever is better.

Next, we tracked the temporal changes in $f_{THz}$ to demonstrate the real-time capability of this method. Figure 5(a) and Media 1 show the temporal changes in $f_{THz}$ when $f_{THz}$ was manually changed within a frequency range of 0.1 THz ± 100 Hz (measurement rate = 100 Hz). A slight change of several Hz to a few tens of Hz in $f_{THz}$ was sensitively reflected in the chart, indicating the potential of this method for precise measurement of absolute frequency in real time.

On the other hand, practical CW-THz sources often exhibit large fluctuations of $f_{THz}$ due to free-running operation or mode hopping. In particular, frequency jumps due to mode hopping are sometimes as high as several hundreds MHz to several GHz. In this case, the CW-THz wave may suddenly cross over many frequency modes of the PC-THz comb because the frequency modes of PC-THz comb are too densely distributed along the frequency axis compared with the frequency change. Such large and/or instantaneous fluctuations of $f_{THz}$ cannot be observed by the conventional method with a single PC-THz comb due to the non-real-time two-step measurement principle used in that method. To evaluate the applicability of our method to such CW-THz waves, we stepwise tuned $f_{THz}$ at intervals of 200 MHz from 99,801,000,000 Hz to 100,440,000,000 Hz. Figure 5(b) shows the result of real-time monitoring of $f_{THz}$ (measurement rate = 100 Hz), indicating that the CW-THz wave crossed two PC-THz comb modes in every stepwise tuning of $f_{THz}$. This demonstration reveals the potential of our method for measuring large and rapid fluctuations of $f_{THz}$.

4.2 Use of dual PC-THz combs without stabilization of frequency spacing

In previous studies, the frequency spacing of the PC-THz comb has been precisely stabilized by using laser control [4, 5, 7]. However, use of a stabilized femtosecond laser has often restricted the use of the THz-comb-referenced spectrum analyzer in various applications, despite its superior performance. If the real-time absolute frequency measurement with dual PC-THz combs could be implemented using free-running, that is, unstabilized, lasers, the scope of application of the spectrum analyzer would be greatly expanded. Recently, temporally serial, two-step frequency measurement has been performed using a single PC-THz comb without stabilization of the frequency spacing [6]; however, there have been no attempts to perform the real-time frequency measurement using dual PC-THz combs without stabilization of the frequency spacing, that is, free-running dual PC-THz combs. We attempted to determine the absolute frequency of a CW-THz wave in real time using free-running dual PC-THz combs.

Consider Eq. (3) when both $f_{rep1}$ and $f_{rep2}$ fluctuate. This relation must always be true every moment regardless of the fluctuations of $f_{rep1}$ and $f_{rep2}$ because a free-running PC-THz comb can also be used as a frequency ruler with a linear scale at every moment. This means that precise stabilization of $f_{rep1}$ and $f_{rep2}$ in dual lasers is not essential if $f_{rep1}$, $f_{rep2}$, $f_{beat1}$, and $f_{beat2}$ are measured simultaneously. In other words, free-running dual PC-THz combs can be used in the proposed method.

First, we evaluated the frequency fluctuation of $mf_{rep1}$ ($m$ = 1,000 and $f_{rep1}$ = 100 MHz) for free-running and stabilized PC-THz combs with respect to various gate times, as shown in Fig. 6. Although the frequency fluctuation of the free-running PC-THz comb was larger than that of the stabilized PC-THz comb, the former was still smaller than the frequency spacing of the PC-THz combs. This means that we could determine the absolute frequency of $f_{THz}$ using the same $m$ value of the PC-THz combs during the measurement.

Next, we measured $f_{rep1}$, $f_{rep2}$, $f_{beat1}$, and $f_{beat2}$ for the frequency-stabilized CW-THz wave, and then determined $f_{THz}$. In this experiment, the absolute frequency of the CW-THz wave was fixed at 100,001,004,000 Hz, whereas $f_{rep1}$ (≈ 100,000,007 Hz) and $f_{rep2}$ (≈ 100,000,217 Hz) were not stabilized. Figure 7 shows the results for the real-time monitoring of (a) $f_{rep1}$, (b) $f_{rep2}$, (c) $f_{beat1}$, (d) $f_{beat2}$, and (e) the corresponding $f_{THz}$ (sampling rate = 10 MHz). Even though $f_{rep1}$ and $f_{rep2}$ were not stabilized, we achieved the frequency precision of 4.0×10$^{-11}$ at a measurement rate of 100 Hz. This precision was comparable to that in the stabilized dual PC-THz combs, indicating that the frequency precision achieved in the proposed method is independent of the frequency stability of $f_{rep1}$ and $f_{rep2}$ (see Fig. 6) if they are precisely measured simultaneously.

Finally, we performed real-time monitoring of $f_{THz}$ which was stepwise tuned at intervals of 200 MHz from 99,801,000,000 Hz to 100,401,000,000 Hz to evaluate the applicability of our method for measuring $f_{THz}$ with large fluctuations. Figure 8 shows the result of real-time monitoring of $f_{THz}$ (measurement rate = 100 Hz), indicating that the CW-THz wave crossed two PC-THz comb modes in every stepwise tuning of $f_{THz}$. Thus, we successfully demonstrated the real-time and precise monitoring of $f_{THz}$ with free-running dual PC-THz combs.

## 5. Conclusions

We demonstrated real-time, precise measurement of the absolute frequency of a CW-THz wave by simultaneous measurement of $f_{rep1}$, $f_{rep2}$, $f_{beat1}$, and $f_{beat2}$ using dual PC-THz combs with different frequency spacings. Regardless of the presence or absence of frequency control of the PC-THz combs, a frequency precision of 10$^{-11}$ was achieved at a measurement rate of 100 Hz. The proposed method was successfully applied to real-time monitoring of $f_{THz}$ with large frequency fluctuations across several PC-THz comb modes, indicating the high potential of our method to practical CW-THz sources with free-running operation or mode hopping. The proposed method will be a practical tool for the characterization and frequency calibration of a variety of CW-THz sources, including THz quantum cascade lasers [17], photomixing sources [18], resonant tunneling diodes [19], and so on.

One may consider that the need for dual femtosecond lasers is still an obstacle for the practical use of this method, even though free-running lasers can be used. Recently, a dual-wavelength mode-locked fiber laser has been realized under certain cavity configurations [20]. Because of dispersion, resulting in different refractive indexes at the two wavelengths in the fiber cavity, the two wavelength lights have different repetition frequencies. This laser will be preferable for the real-time absolute frequency measurement based on dual PC-THz combs with different frequency spacings. Another possible method is to use a single, free-running femtosecond laser with frequency modulation of $f_{rep}$. Work is in progress to perform real-time monitoring of $f_{THz}$ with a single, $f_{rep}$-modulated femtosecond laser. The proposed method, in combination with these lasers, will further allow the practical use of THz-comb-referenced spectrum analyzers, and will hence accelerate their adoption in real-world applications.

Acknowledgment

This work was supported by Collaborative Research Based on Industrial Demand

from Japan Science and Technology Agency, Grants-in-Aid for Scientific Research No. 26246031 from the Ministry of Education, Culture, Sports, Science, and Technology of Japan.

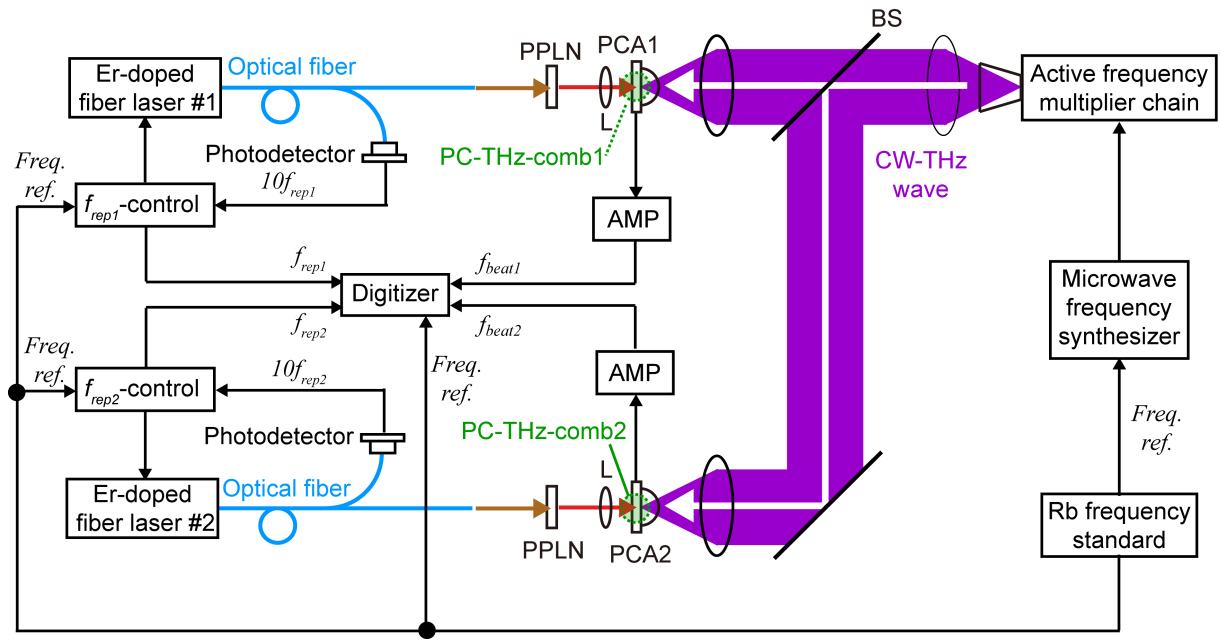

Fig. 1. Principle of real-time absolute frequency measurement of CW-THz wave based on simultaneous measurement of $f_{beat1}$, $f_{beat2}$, $f_{rep1}$, and $f_{rep2}$ with dual PC-THz combs having different frequency spacings.

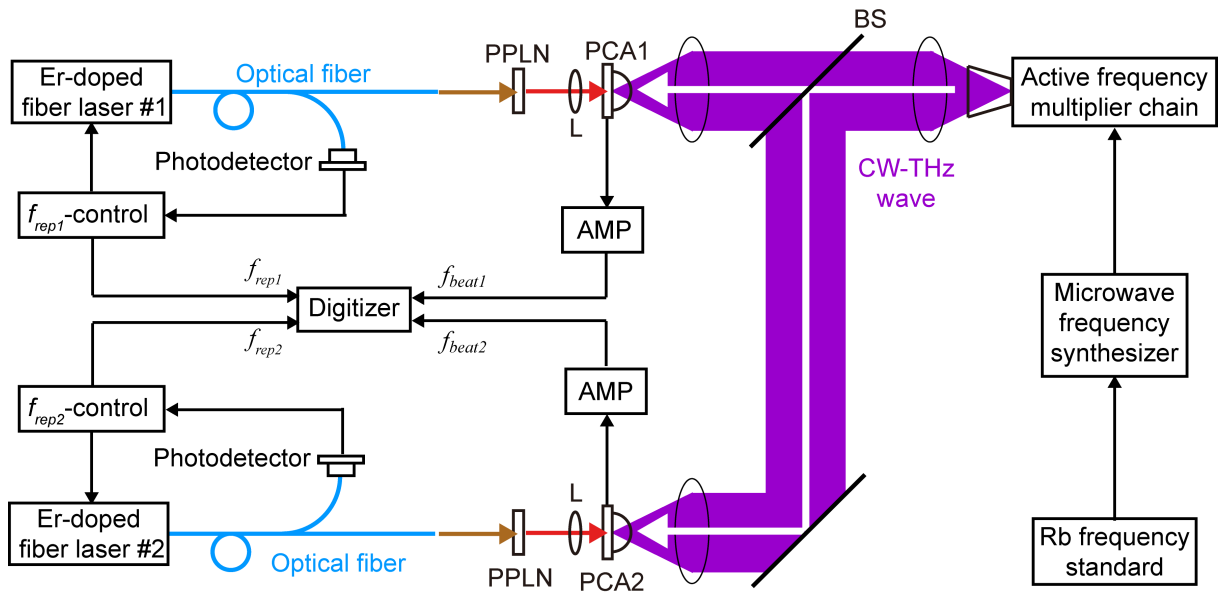

Fig. 2. Experimental setup. PPLN: periodically-poled-lithium-niobate crystal; BS: beam splitter; L: objective lens; PCA1 and PCA2: photoconductive antennas; AMP: current preamplifier.

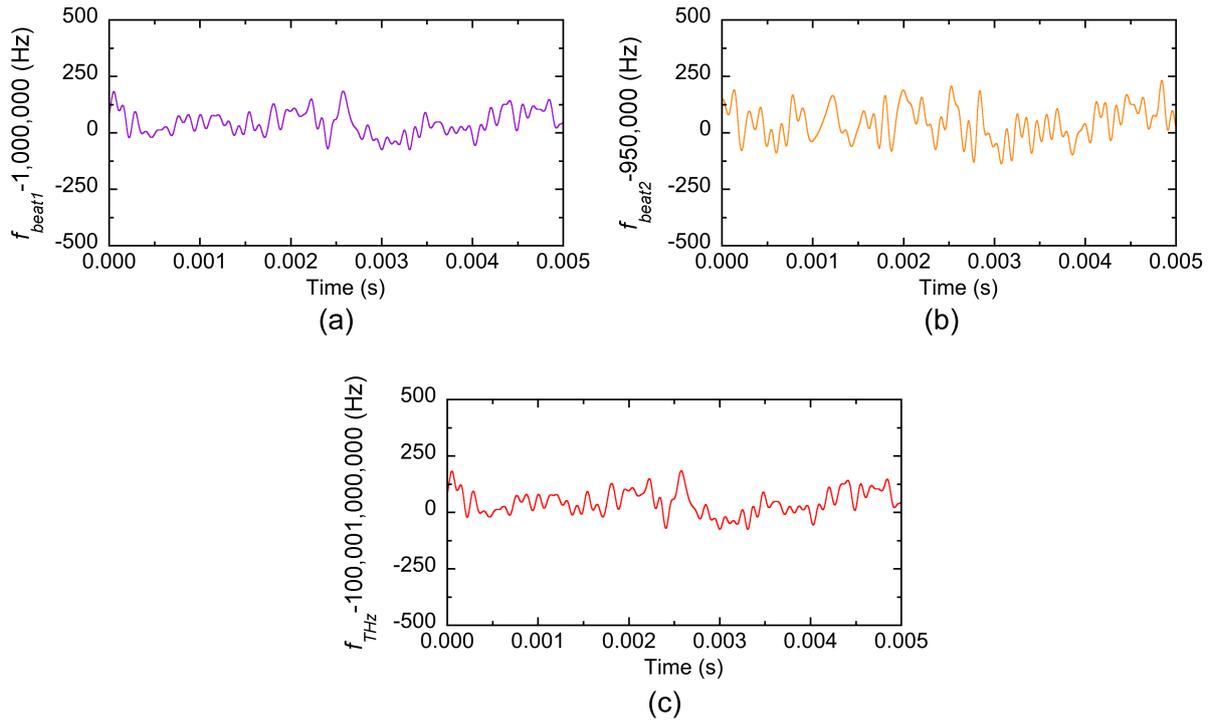

Fig. 3. Temporal changes of (a) $f_{beat1}$, (b) $f_{beat2}$, and (c) $f_{THz}$ when $f_{rep1}$ and $f_{rep2}$ were fixed at 100,000,000 Hz and 100,000,050 Hz by laser stabilization control. Sampling rate was 10 MHz.

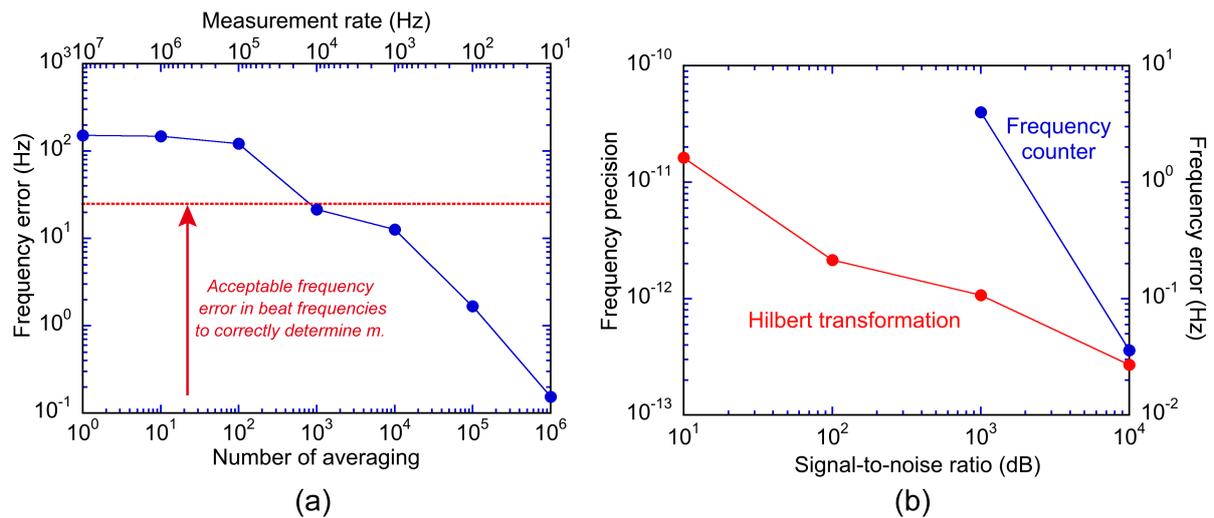

Fig. 4. (a) Frequency error in $f_{beat1}$ with respect to number of averaged signals or measurement rate. (b) Comparison of frequency precision and the corresponding frequency error in $f_{THz}$ with respect to signal-to-noise ratio between Hilbert transformation method and frequency counter method. The measurement rate was set at 10 Hz.

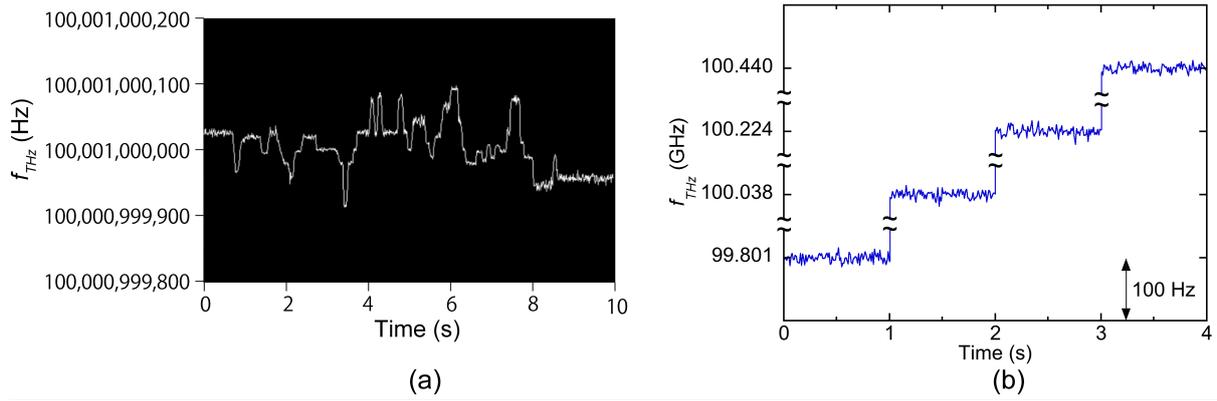

Fig. 5. (a) Temporal change of $f_{THz}$ when the $f_{THz}$ was manually changed within a frequency range of 0.1 THz ± 100 Hz. (b) Temporal change of $f_{THz}$ when $f_{THz}$ was stepwise tuned at intervals of 200 MHz from 99,801,000,000 Hz to 100,440,000,000 Hz. The measurement rate was set at 100 Hz for both measurements.

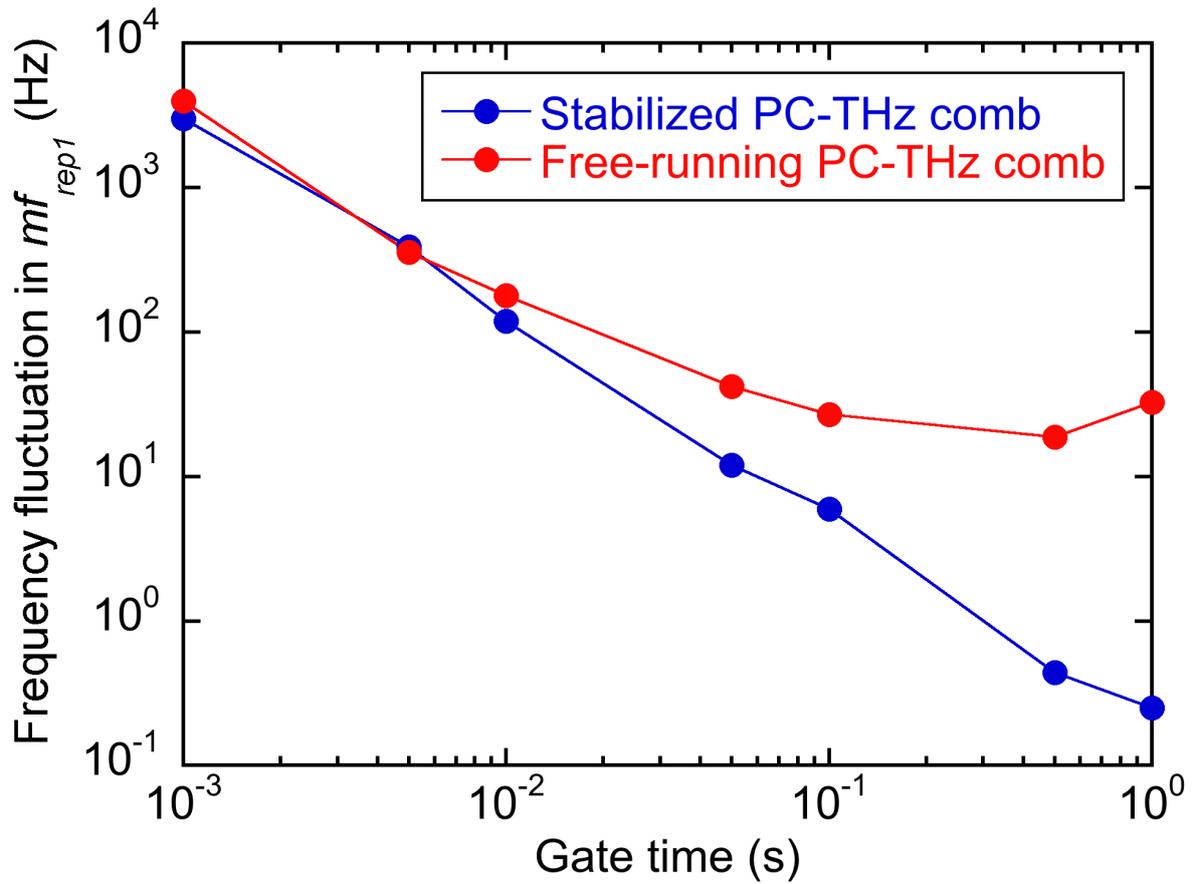

Fig. 6. Comparison of frequency fluctuation in $mf_{rep}$ with respect to gate time between stabilization control and free-running operation.

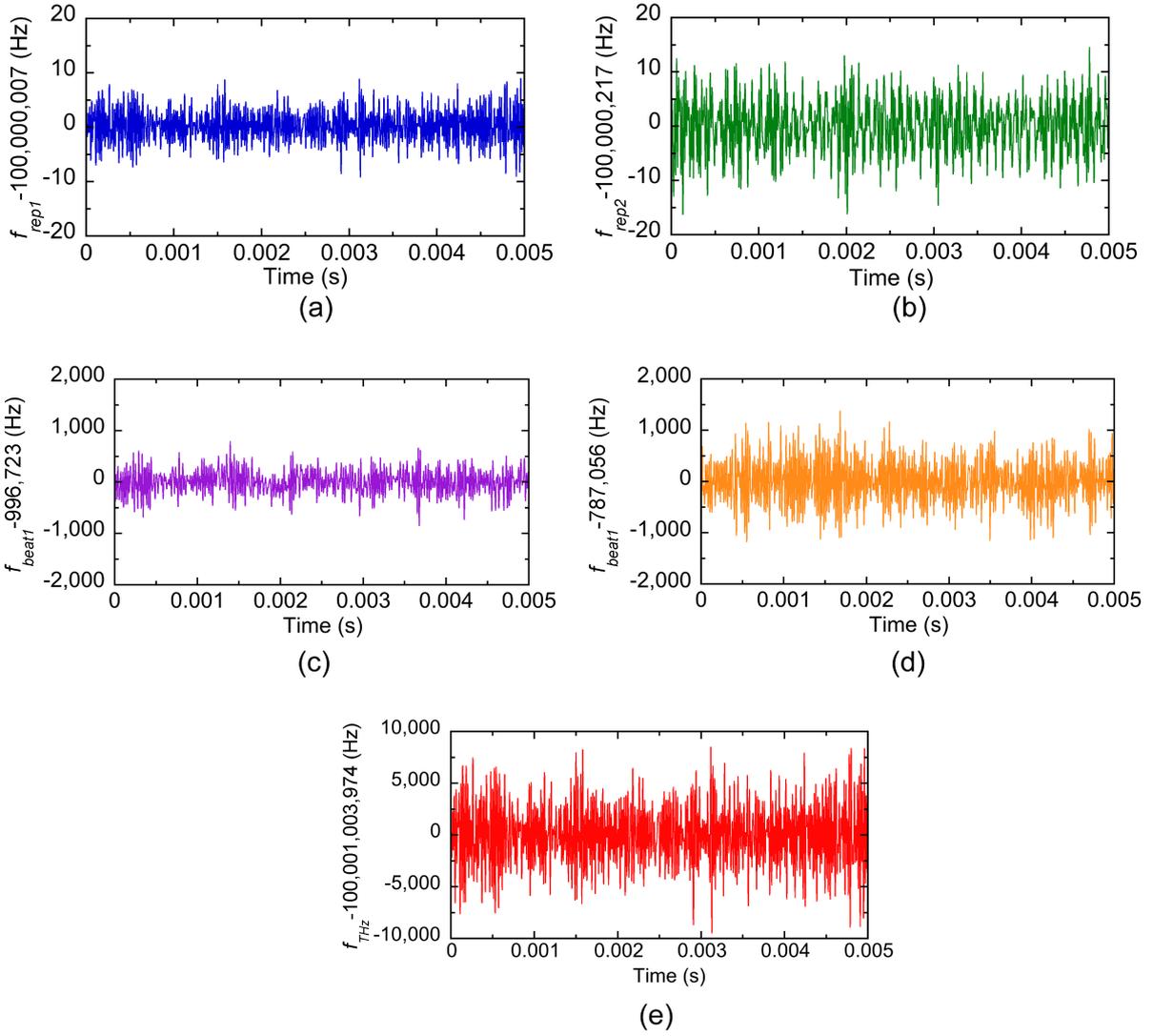

Fig. 7. Temporal changes of (a) $f_{beat1}$, (b) $f_{beat2}$, (c) $f_{rep1}$, (d) $f_{rep2}$, and (e) $f_{THz}$ when $f_{rep1}$ and $f_{rep2}$ were not stabilized. Sampling rate was 10 MHz.

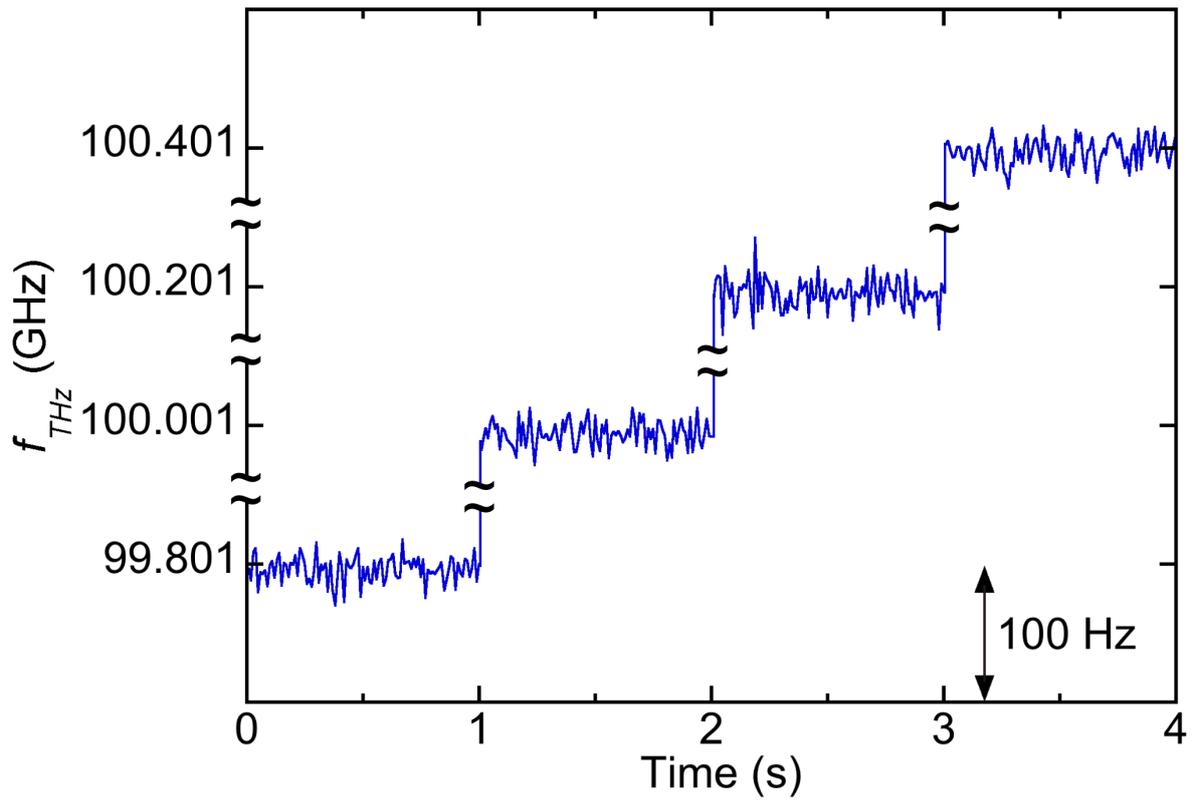

Fig. 8. Temporal change of $f_{THz}$ when $f_{THz}$ was stepwise tuned at intervals of 200 MHz from 99,801,000,000 Hz to 100,401,000,000 Hz. The measurement rate was set at 100 Hz.